\documentclass[10pt,letterpaper]{article}
\usepackage[top=0.85in,left=2.75in,footskip=0.75in,marginparwidth=2in]{geometry}

\usepackage[utf8]{inputenc}
\DeclareUnicodeCharacter{221A}{$\surd$}

\usepackage{cite}

\usepackage{nameref,hyperref}

\usepackage[right]{lineno}

\usepackage{microtype}

\raggedright
\usepackage{changepage}

\usepackage[aboveskip=1pt,labelfont=bf,labelsep=period,singlelinecheck=off]{caption}

\usepackage{subcaption}
\newcommand{\psubref}[1]{(\subref{#1})} %

\usepackage{booktabs}

\makeatletter
\renewcommand{\@biblabel}[1]{\quad#1.}
\makeatother

\usepackage{lastpage,fancyhdr,graphicx}

\graphicspath{{figures/}} %
\usepackage{epstopdf}
\fancyheadoffset[L]{2.25in}
\fancyfootoffset[L]{2.25in}

\usepackage{color}

\definecolor{Gray}{gray}{.25}

\usepackage{graphicx}

\usepackage{sidecap}

\usepackage{wrapfig}
\usepackage[pscoord]{eso-pic}
\usepackage[fulladjust]{marginnote}
\reversemarginpar

\usepackage{gensymb}

\usepackage[numbers,sort&compress]{natbib}

\begin{document}
\vspace*{0.35in}

\begin{flushleft}
{\Large
  \textbf\newline{
    Urban climate regressions require representative monitoring networks
  }
}
  \\
  Mart\'i Bosch\textsuperscript{1,2},
  Aldo Brandi\textsuperscript{3},
  Christoph Bachofen\textsuperscript{4,5},
  Perrine Hamel\textsuperscript{6},  
  Roy P. Remme\textsuperscript{7},
  Emma E. Ramsay\textsuperscript{6},
  Gabriele Manoli\textsuperscript{3},  
  Martin A. Schlaepfer\textsuperscript{8},
  Moritz Burger\textsuperscript{9,10}
  \\
  \bigskip
  {\small
    \textsuperscript{1} Center for Climate Systems Modeling, Eidgenössische Technische Hochschule Zürich, Zürich, Switzerland \\
    \textsuperscript{2} Federal Office of Meteorology and Climatology MeteoSwiss, Zürich, Switzerland \\
    \textsuperscript{3} Laboratory of Urban and Environmental Systems, École Polytechnique Fédérale de Lausanne, Lausanne, Switzerland \\
    \textsuperscript{4} Plant Ecology Research Laboratory, École Polytechnique Fédérale de Lausanne, Lausanne, Switzerland \\
    \textsuperscript{5} Functional Plant Ecology, Community Ecology Unit, Swiss Federal Institute for Forest, Snow and Landscape WSL, Lausanne, Switzerland \\
    \textsuperscript{6} Asian School of the Environment, Nanyang Technological University, Singapore, Singapore \\
    \textsuperscript{7} Institute of Environmental Sciences, Leiden University, Leiden, The Netherlands \\    
    \textsuperscript{8} Institute for Environmental Sciences, University of Geneva, Geneva, Switzerland \\
    \textsuperscript{9} Oeschger Centre for Climate Change Research, University of Bern, Bern, Switzerland \\
    \textsuperscript{10} Institute of Geography, University of Bern, Bern, Switzerland
    \\
    \bigskip
    * Corresponding authors: boschma@ethz.ch (M.B.), aldo.brandi@epfl.ch (A.B.)
}
\newline

\end{flushleft}

\section*{Abstract}

The increasing frequency of heatwaves in urban areas threatens public health and calls for effective mitigation and adaptation strategies.  %
Land use regression models relate intra-urban air temperature to spatial predictors such as building volume, tree canopy cover and topography, and are increasingly used to produce fine-scale temperature maps. However, the accuracy of these estimates can be compromised by unrepresentative monitoring networks.
Here we show that reference weather station networks, designed for regional rather than intra-urban monitoring, sample only a narrow portion of a city's spatial heterogeneity, missing both the densely built core and the cool, vegetated periphery that set the two ends of the urban temperature range.
As a result, city-wide predictions underestimate temperature differences, in ways that cross-validation against the station measurements does not detect.
Using Zurich, Switzerland, as a case study, we compare multiple linear regression, generalized additive models and random forests as up to 117 low-cost temperature sensors are added to the four reference weather stations.
When trained on reference stations alone, all three models achieve comparable cross-validated performance ($R^2_{\mathrm{CV}} \approx 0.79$, $\mathrm{RMSE}_{\mathrm{CV}} \approx 0.70$\,\degree C).
As low-cost sensors broaden the sampled spatial heterogeneity, multiple linear regression scores lower on both metrics ($R^2_{\mathrm{CV}} = 0.57$, $\mathrm{RMSE}_{\mathrm{CV}} = 1.00$\,\degree C), whereas the non-linear models remain at their reference-only level ($R^2_{\mathrm{CV}} = 0.79$, $\mathrm{RMSE}_{\mathrm{CV}} = 0.69$\,\degree C).
Although a third of the low-cost sensors already recovers nearly the full predicted temperature range, which predictors are identified, and with which sign, is determined by the station sample rather than the model class.
At full coverage, only multiple linear regression leaves significant residual spatial autocorrelation, suggesting a linear form unable to capture non-linear responses of air temperature to urban form.
Drawing on landscape ecology research on focal site study design, we provide a transferable framework for diagnosing station network representativeness and improving the reliability of city-wide environmental maps.

\section*{Introduction}

The urban fabric alters local climatic characteristics, often making built-up areas warmer than their rural surroundings --- a phenomenon known as the urban heat island (UHI) effect \cite{oke1982energetic}. Heat extremes in urban areas have been associated with numerous negative impacts, including increased water and energy consumption, reduced labour productivity and increased risk of heat-related morbidity and mortality \cite{guhathakurta2007impact,phelan2015urban,santamouris2015impact,li2019urban,sera2019urban,casanueva2020escalating,santamouris2020recent,masselot2023excess,ragettli2023explorative}. Given that cities concentrate a growing share of the global population \cite{un2025world} and are increasingly exposed to rising temperatures and heat waves \cite{oleson2011examination,chapman2017impact,zhao2021global,li2024persistent}, mitigating urban heat has become a central planning priority, which requires understanding how the built environment and green infrastructure influence the local microclimate \cite{deilami2018urban,rosenzweig2018climate,geneletti2020planning,masson2020urban}.

Land use regression (LUR) models have emerged as a widely used tool for characterizing the relationship between environmental variables such as air pollution or temperature and spatial predictors of land use and urban form \cite{jerrett2005review,hoek2008review}.
When applied to urban temperature, LUR models relate air temperature measurements at monitoring stations to spatially explicit predictors --- such as building volume, tree canopy cover, topographical features or distance to water bodies --- enabling fine-scale thermal mapping and the quantification of each predictor's contribution to local temperature variability \cite{zhou20142011,shi2018modelling,shaker2019investigating,shi2019assessing,tsin2020land,burger2021modelling}.
Initially based on multiple linear regression, LUR increasingly also draws on more flexible approaches, such as generalized additive models, random forests, and gradient boosting, which can capture non-linear predictor--temperature relationships \cite{grunwald2020predicting,han2022prediction,oukawa2022fine,wang2023machine,huang2026revealing}.
However, a LUR model's reliability and ability to generalize depend critically on whether the monitoring stations represent the full range of urban environments present in a given study area --- a long-standing concern in landscape ecology, where sampling that fails to span the full environmental gradient can bias the inferred relationships between spatial pattern and ecological process \cite{miller2004spatial,meyer2021predicting,gutzwiller2023using}.
Reference automatic weather station (AWS) networks are typically designed for synoptic and regional monitoring rather than for characterizing intra-urban thermal heterogeneity \cite{grimmond2006progress,oke2006initial,grimmond2010climate,world2024guide} and often fail to capture the spatial diversity of urban thermal environments \cite{muller2013sensors,yang2019designing,hamdi2020state,chen2022urban}.

In recent years, dense networks of low-cost temperature sensors (LCS) have proliferated in many cities, offering the potential to capture intra-urban temperature variability at unprecedented spatial resolution \cite{muller2013sensors,hamdi2020state,masson2020urban,wong2025government}.
Although the reliability of such sensors has been questioned, prior work shows they can yield more representative estimates of urban temperature than reference stations sited outside the urban core \cite{bosch2026revisiting}.
In any case, such networks must provide representative coverage of the urban thermal environment to ensure that the resulting LUR models remain reliable across the whole city, yet this is rarely evaluated beforehand.
Landscape ecology confronts this problem in focal site multi-scale study designs \cite{brennen2002focal}, in which an ecological response is measured at a sample of locations and related to spatial predictors computed at multiple buffer scales around each site.
Decades of work in this field have shown that biased site selection can reduce statistical power, produce contradictory findings in the presence of non-monotonic effects, and lead to erroneous identification of relevant predictors when landscape variables are correlated \cite{smith2009confronting,albert2010sampling,eigenbrod2011sub,dormann2013collinearity,pasher2013optimizing,miguet2016determines,hesselbarth2024computational}.
For example, in a survey of more than 1{,}100 sites, whether forest had a positive or negative effect on frog abundance depended on which part of the forest-cover gradient was sampled, so that studies covering different ranges reached opposite conclusions \cite{eigenbrod2011sub}.
The focal site setup is closely analogous to LUR, so the same pitfalls arise: if the stations cover only part of the range of urban form, the predictors identified and the direction of their effects may reflect the station sample rather than the physical mechanisms.

Here we evaluate how AWS and LCS networks in Zurich, Switzerland, cover the spatial heterogeneity of the urban environment, and how this coverage affects LUR model performance and city-wide spatial predictions.
Zurich is a particularly well-suited case study because of both its dense LCS network and its wide environmental gradient, which spans from a densely built core to forested hills.
We progressively augment the four AWS with up to 117 LCS and compare multiple linear regression (MLR), generalized additive model (GAM) and random forest (RF) approaches.
Drawing on the landscape ecology framework of focal-site study design, we interpret the design, evaluation, and predictive use of urban monitoring networks, a connection that, to our knowledge, has not previously been made in this field.
Our results provide operational guidance for the evaluation and design of urban temperature monitoring networks.

\section*{Results}

\subsection*{Reference stations undersample both ends of the urban environmental gradient, low-cost sensors only one}

\begin{figure}
  \captionsetup[subfigure]{justification=centering}
  \begin{adjustwidth}{-.4\textwidth}{0cm}
    \begin{subfigure}[b]{.65\linewidth}
      \includegraphics[width=\linewidth]{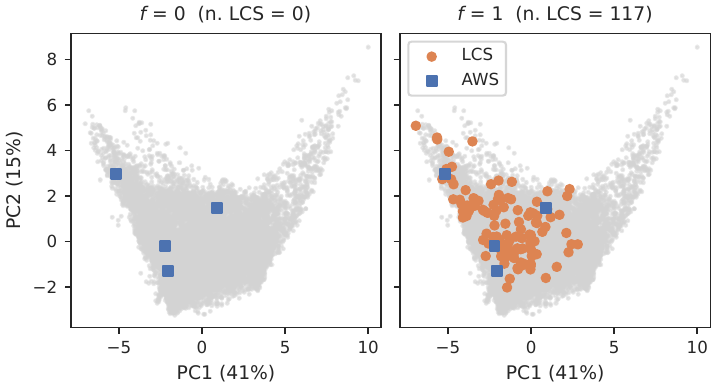}
      \caption{}
      \label{fig:pca-scatter}
    \end{subfigure}%
    \hfill%
    \begin{subfigure}[b]{.34\linewidth}
      \includegraphics[width=\linewidth]{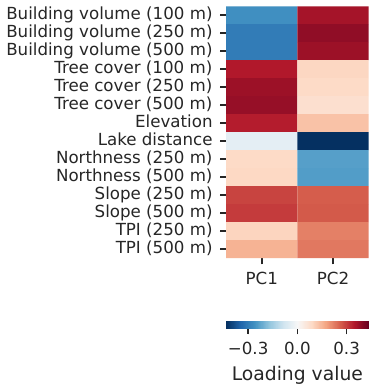}
      \caption{}
      \label{fig:pca-loadings}
    \end{subfigure}
    \begin{minipage}{\linewidth}
      \caption{Station coverage of the predictor feature space according to the PCA. For the first two components, we show \psubref{fig:pca-scatter} the scores of the AWS, LCS and grid cells, and  \psubref{fig:pca-loadings} the loadings of the predictor variables. \label{fig:station-features}}
    \end{minipage}
  \end{adjustwidth}
\end{figure}

The four AWS cover a narrow subset within the city's feature space (\autoref{fig:pca-scatter}), here represented by the spatial predictors computed for every cell of a regular 100\,m grid covering the study area and summarized by a principal component analysis (PCA).
The first axis of the PCA (PC1, explaining 40.9\% of the variance) separates forested, elevated and topographically complex terrain from densely built-up areas.
The second axis (PC2, 14.8\% of the variance) mainly reflects proximity to the lake and building volume, together with slope, northness and topographic position, running from inland, north-facing valley bottoms to densely built, south-facing and convex ground near the lake shore (\autoref{fig:pca-loadings}).
Together the two axes capture 55.7\% of the total variance.

Two distinct regions of feature space are left unsampled by the AWS network.
The first lies beyond the reach of both networks, at the vegetated, elevated end of PC1: 32.2\% of grid cells fall beyond the highest PC1 value of any AWS and 15.2\% beyond that of any LCS station (\autoref{fig:pca-scatter}).
The second comprises environments with low PC1 and high PC2 (densely-built, lakeshore neighbourhoods on convex ground), entirely absent from the AWS sample yet reached by eight LCS stations.
These two gaps likely correspond to the thermal extremes of Zurich's urban landscape --- the cool, vegetated periphery and the intensely built core --- and together mean that any model trained on AWS alone must extrapolate well beyond its training support across both ends of the city's environmental gradient.
While the LCS network fills the densely-built extreme left unsampled by the AWS, it barely extends into the vegetated and high elevation extreme. %
Therefore, even with the full LCS network, no model can be fully validated against the environments that presumably anchor the cool end of the urban temperature range.

\subsection*{Models trained on reference stations alone underestimate the intra-urban temperature range}

\begin{figure}
  \begin{adjustwidth}{-.4\textwidth}{0cm}
    \includegraphics[width=\linewidth]{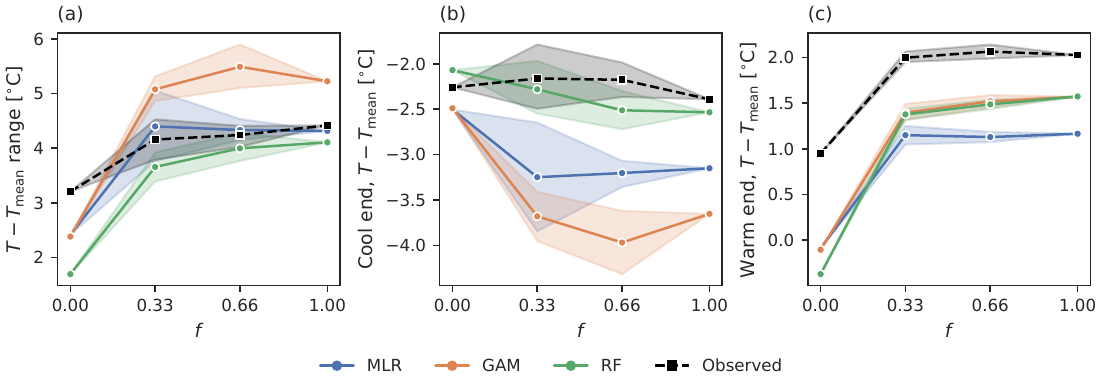}
    {\phantomsubcaption\label{fig:pred-spread-range}}
    {\phantomsubcaption\label{fig:pred-spread-cool}}
    {\phantomsubcaption\label{fig:pred-spread-warm}}
    \begin{minipage}{\linewidth}
      \centering
      \caption{
        Spread of the predicted $T - T_{\mathrm{mean}}$ over the 100\,m prediction grid as a function of the LCS fraction $f$: \psubref{fig:pred-spread-range} range between the 5th and 95th percentiles across grid cells, \psubref{fig:pred-spread-cool} cool (5th percentile) and \psubref{fig:pred-spread-warm} warm (95th percentile) ends.  %
        Markers show the mean over the five subsampling repetitions and shaded bands bootstrap 95\% confidence intervals, which collapse at $f=0$ and $f=1$, where every repetition fits the same stations.
        The dashed line shows the same quantities among the stations each model was trained on.  %
      }
      \label{fig:pred-spread}
    \end{minipage}
  \end{adjustwidth}
\end{figure}

In order to represent the intra-urban distribution of temperatures, we train models to predict the temperature anomaly $T - T_{\mathrm{mean}}$, i.e., how much warmer or cooler each location is than the station mean at the same hour (see \nameref{sec:air-temperature-data}).
To assess the role of network coverage, we augment the four AWS with a growing fraction $f$ of the LCS network, from none ($f=0$) to 39, 77 and all 117 LCS stations ($f = 0.33$, $0.66$ and $1$), with five random subsamples at the intermediate fractions, and compare the models fitted on each configuration (see \nameref{sec:lur-setup}).
Among the stations, the observed range of $T - T_{\mathrm{mean}}$ grows from $3.2$\,\degree C for the four AWS to $4.2$\,\degree C at $f=0.33$ and $4.4$\,\degree C with the full network (\autoref{fig:pred-spread-range}).
The models trained on the four AWS predict narrower ranges than the observed $3.2$\,\degree C, i.e., $2.4$, $2.4$ and $1.7$\,\degree C for MLR, GAM and RF respectively.
This narrower range does not arise from the cool end, which all three models reproduce to within about $0.3$\,\degree C (\autoref{fig:pred-spread-cool}), but from the warm end, which they predict from $1.0$\,\degree C (MLR) to $1.3$\,\degree C (RF) below that of the AWS (\autoref{fig:pred-spread-warm}).
Trained with the full network, the models diverge: RF still predicts a narrower range than observed ($4.1$ against $4.4$\,\degree C), MLR roughly matches it ($4.3$\,\degree C), and GAM exceeds it ($5.2$\,\degree C).
Most of this widening is already achieved with a third of the LCS network, at which point all predicted ranges are very close to their corresponding full-network values (\autoref{fig:pred-spread-range}).
The variability across subsampling repetitions is also largest at this fraction, showing that which stations are added matters most while the network is still small.

\begin{figure}
  \begin{adjustwidth}{-.4\textwidth}{0cm}
    \includegraphics[width=\linewidth]{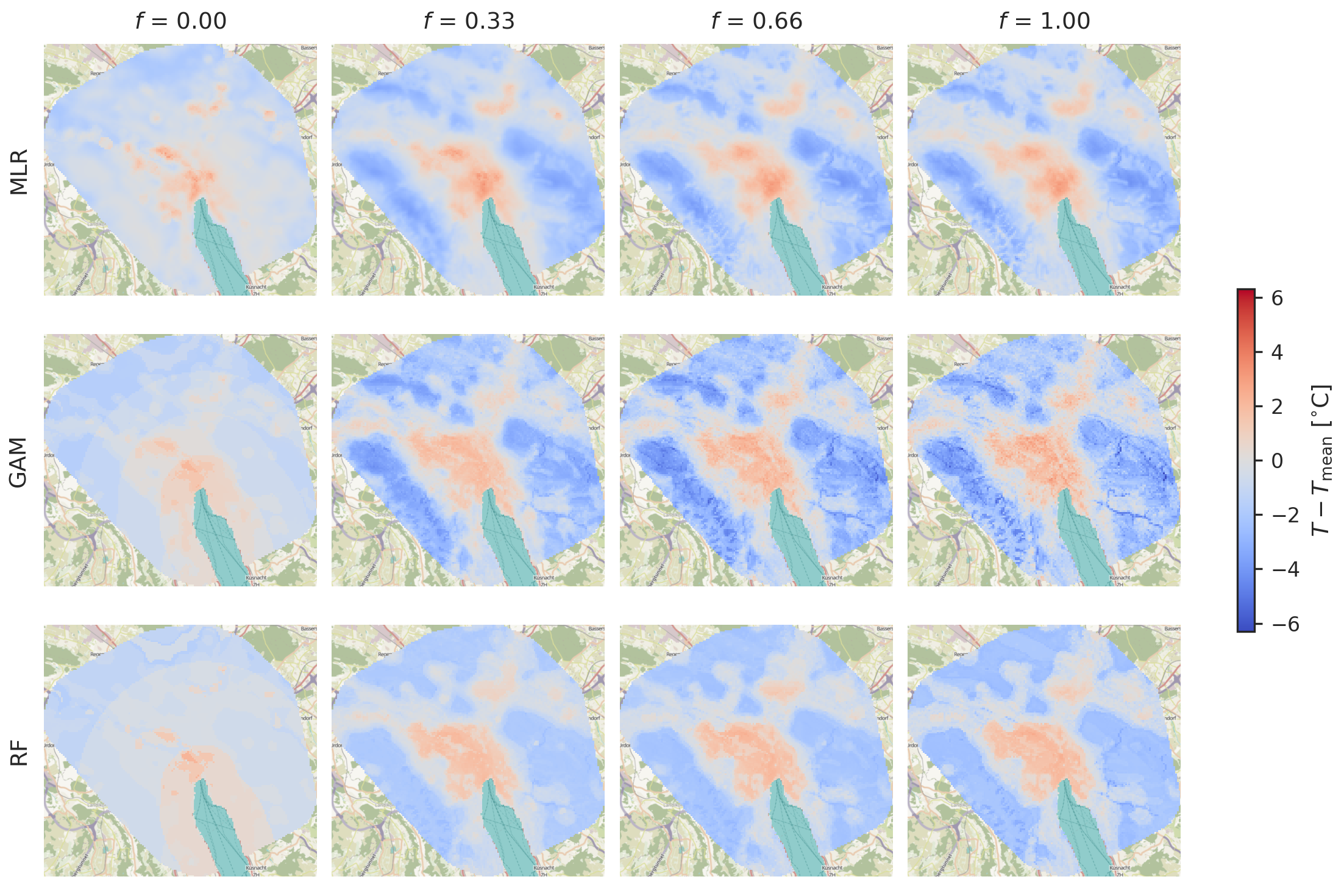}
    \begin{minipage}{\linewidth}
      \centering
      \caption{Predicted $T - T_{\mathrm{mean}}$ maps for the fitted models across LCS fractions $f$ (0, 39, 77, and 117 stations), each relative to the mean of the stations used to fit the model. For the intermediate fractions ($f = 0.33$, $0.66$), each panel shows the per-cell mean across the five subsampling repetitions; the $f=0$ and $f=1$ cases are deterministic. All maps use a common color scale set to the absolute maximum prediction across all model--fraction combinations. The corresponding per-cell standard deviation across repetitions, quantifying the sensitivity to the specific LCS subset drawn, is mapped in the section ``Grid prediction maps'' of the results notebook (\nameref{sec:code-lur-results}).}
      \label{fig:regr-maps}
    \end{minipage}
  \end{adjustwidth}
\end{figure}

The narrower temperature differences of the AWS-only models are visible in the city-wide maps (\autoref{fig:regr-maps}), which consistently show a warm anomaly around the city center but with a magnitude that clearly varies across fractions and models.
With AWS alone, the warm core and the spatial contrasts around it are less pronounced, whereas from $f=0.33$ onwards, the warm anomaly intensifies and cool bands emerge along the forested hills around the city, most markedly for GAM and least for RF.
Beyond this fraction, adding further LCS changes the maps only slightly.

\begin{figure}
  \begin{adjustwidth}{-.4\textwidth}{0cm}
    \includegraphics[width=\linewidth]{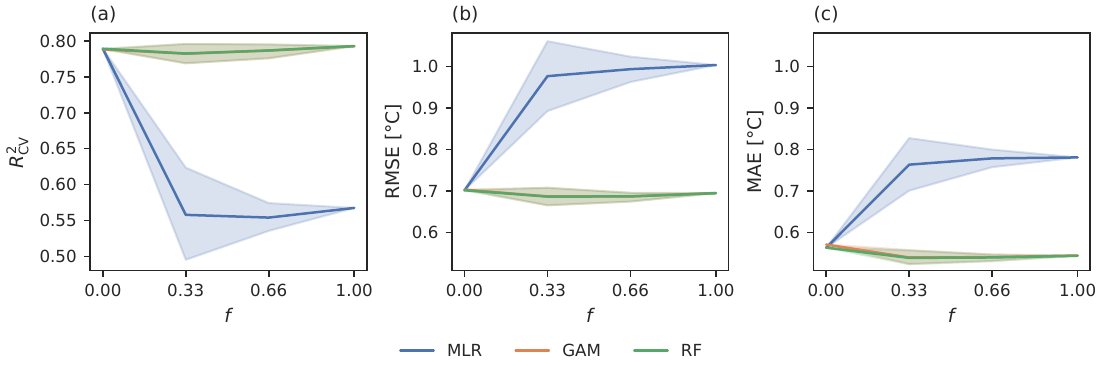}
    {\phantomsubcaption\label{fig:regr-scores-r2}}
    {\phantomsubcaption\label{fig:regr-scores-rmse}}
    {\phantomsubcaption\label{fig:regr-scores-mae}}
    \begin{minipage}{\linewidth}
      \centering
      \caption{Cross-validated model performance across LCS fractions $f$ for MLR, GAM and RF: \psubref{fig:regr-scores-r2} $R^2_{\mathrm{CV}}$, \psubref{fig:regr-scores-rmse} $\mathrm{RMSE}_{\mathrm{CV}}$ and \psubref{fig:regr-scores-mae} $\mathrm{MAE}_{\mathrm{CV}}$, the latter two sharing a common $y$-axis. Shaded bands show 95\% confidence intervals across the five repetitions; at $f = 0$ and $f = 1$ the sample is deterministic.}
      \label{fig:regr-scores}
    \end{minipage}
  \end{adjustwidth}
\end{figure}

The narrower temperature range of the AWS-only models and the weaker spatial structure of their maps both go unnoticed in cross-validation.
These models achieve $R^2_{\mathrm{CV}} \approx 0.79$ and $\mathrm{RMSE}_{\mathrm{CV}} \approx 0.70$\,\degree C (\autoref{fig:regr-scores}), which would conventionally be read as a good fit.
As LCS are added, cross-validation reveals different behavior among model classes: MLR scores markedly lower ($R^2_{\mathrm{CV}} \approx 0.57$, $\mathrm{RMSE}_{\mathrm{CV}} \approx 1.0$\,\degree C), whereas GAM and RF keep their scores, likely due to their greater flexibility in fitting the conditions at each station.
Yet with the full network, GAM predicts a wider temperature range than observed at the stations and RF a narrower one, even though both achieve the same cross-validated scores.  %
Good cross-validated performance therefore does not guarantee reliable spatial prediction: evaluated at the station sites only, it reflects how homogeneous the training sample is and cannot tell how a model extrapolates beyond those locations.
Nor do cross-validation scores reveal whether the remaining errors are spatially structured, which we examine next.

\subsection*{Buffer overlap does not increase residual spatial autocorrelation}

\begin{figure}
  \begin{adjustwidth}{-.4\textwidth}{0cm}
    \includegraphics[width=\linewidth]{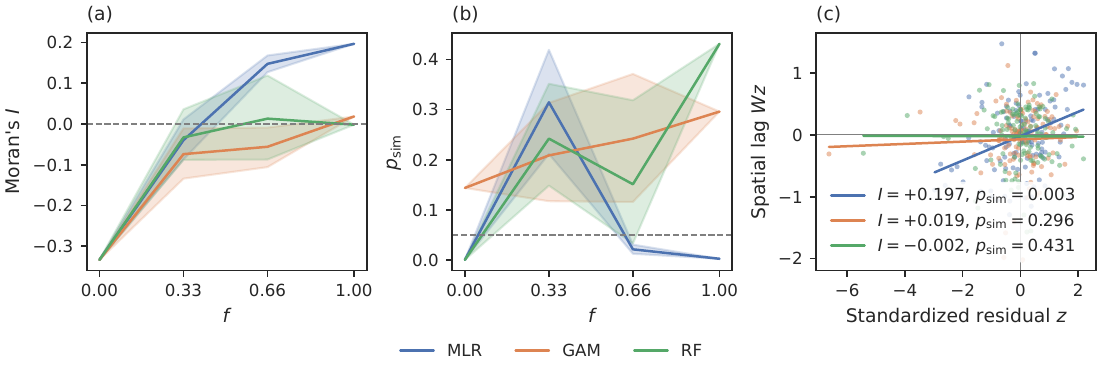}
    {\phantomsubcaption\label{fig:regr-rsa-moran}}
    {\phantomsubcaption\label{fig:regr-rsa-psim}}
    {\phantomsubcaption\label{fig:regr-rsa-scatter}}
    \begin{minipage}{\linewidth}
      \centering
      \caption{Residual spatial autocorrelation of the out-of-fold residuals for MLR, GAM and RF. \psubref{fig:regr-rsa-moran} Residual Moran's $I$ and \psubref{fig:regr-rsa-psim} its permutation $p$-value ($p_{\mathrm{sim}}$) across LCS fractions $f$, with dashed reference lines at $I = 0$ and $\alpha = 0.05$; shaded bands show 95\% confidence intervals across the five repetitions (at $f = 0$ and $f = 1$ the sample is deterministic). \psubref{fig:regr-rsa-scatter} Moran scatterplot at $f=1$, showing each station's standardized residual $z$ against the mean residual of its four nearest neighbours ($Wz$); the slope of each fitted line equals Moran's $I$.}
      \label{fig:regr-rsa}
    \end{minipage}
  \end{adjustwidth}
\end{figure}

Regression models assume independent errors, an assumption that is often violated with spatial data, where nearby observations tend to be similar.
Residual spatial autocorrelation, i.e., a correlation between the prediction errors of neighbouring stations, is the standard diagnostic for such violations.
A frequently raised source of such dependence in focal-site designs is buffer overlap: the areas over which the predictors of two nearby stations are computed can overlap, so that their predictor values, and potentially their errors, are not independent \cite{zuckerberg2012overlapping,zuckerberg2020review}.
Were buffer overlap to violate the independence assumption, densely-spaced stations should produce residuals with significant positive spatial autocorrelation.
We therefore track residual Moran's $I$ as the LCS network, and hence the degree of buffer overlap, progressively densifies.
At $f=0$, with only $n=4$ stations forming a fully-connected $k=3$ nearest-neighbour graph, Moran's $I$ equals the expected null $I=-1/(n-1)$ for any residual vector and the permutation test is uninformative (\autoref{fig:regr-rsa}).
Once a third of the LCS network is added ($f=0.33$), Moran's $I$ collapses from this degenerate null toward zero for all three models (\autoref{fig:regr-rsa-moran}).
For GAM and RF, it stays there as coverage grows: across $f = 0.33$, $0.66$ and $1$ their residual autocorrelation remains within $[-0.074, +0.019]$ and non-significant (mean $p_{\mathrm{sim}} \geq 0.151$; \autoref{fig:regr-rsa-psim}), even as the number of LCS, and with it buffer overlap, increases from 39 to 117.
In contrast, MLR drifts towards positive values, becoming significant once two thirds of the network are added ($I = +0.148$, $p_{\mathrm{sim}} = 0.022$ at $f=0.66$).  %

At $f=1$, residual Moran's $I$ is close to zero for the non-linear models ($I_{\mathrm{GAM}} = +0.019$, $p_{\mathrm{sim}} = 0.296$; $I_{\mathrm{RF}} = -0.002$, $p_{\mathrm{sim}} = 0.431$; \autoref{fig:regr-rsa-scatter}), indicating that the spatial structure of the temperature anomaly is largely captured once the full station network is used, whereas MLR retains a modest but significant positive autocorrelation ($I_{\mathrm{MLR}} = +0.197$, $p_{\mathrm{sim}} = 0.003$).
This is consistent with the lower cross-validated performance of MLR: a linear model that cannot represent non-linear responses leaves systematic errors wherever predictors take similar values, and since nearby stations tend to share similar surroundings, these errors cluster in space.
Why residual autocorrelation remains non-significant for the non-linear models despite the dense buffer overlap of the LCS network is examined in the discussion.

\subsection*{Network composition shapes predictor selection and dominant buffer scale}

\begin{figure}
  \begin{adjustwidth}{-.4\textwidth}{0cm}
    \centering
    \begin{subfigure}[t]{.473\linewidth}
      \phantomcaption\label{fig:importance-heatmap}
      \begin{minipage}[t]{1.5em}\vspace{0pt}\psubref{fig:importance-heatmap}\end{minipage}%
      \begin{minipage}[t]{\dimexpr\linewidth-1.5em\relax}\vspace{0pt}%
        \vspace{.094\linewidth}%
        \includegraphics[width=\linewidth]{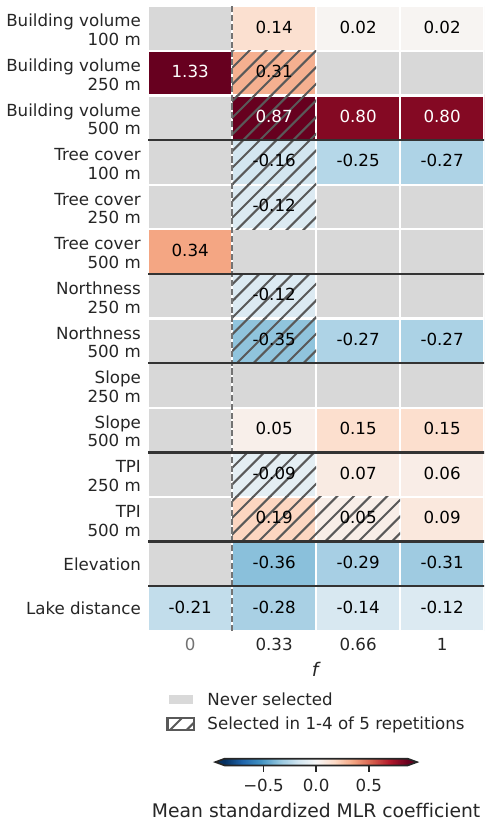}%
      \end{minipage}
    \end{subfigure}%
    \hfill%
    \begin{subfigure}[t]{.507\linewidth}
      \phantomcaption\label{fig:importance-headline}
      \begin{minipage}[t]{1.5em}\vspace{0pt}\psubref{fig:importance-headline}\end{minipage}%
      \begin{minipage}[t]{\dimexpr\linewidth-1.5em\relax}\vspace{0pt}%
        \includegraphics[width=\linewidth]{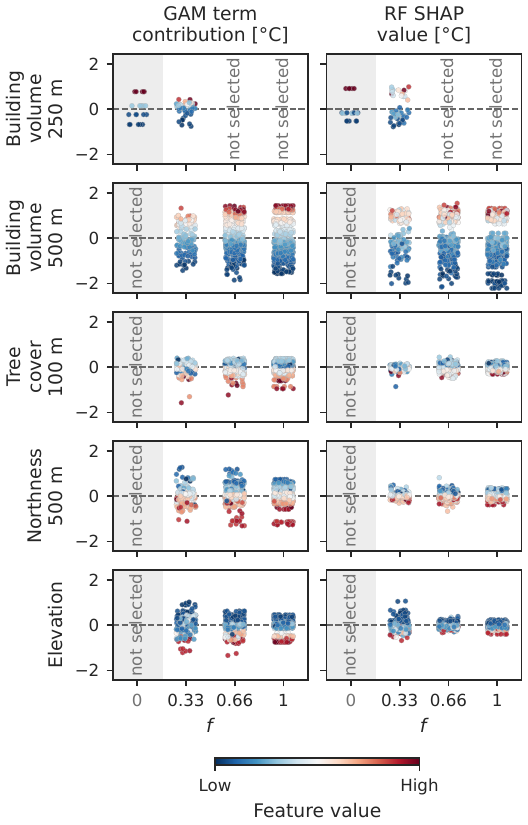}%
      \end{minipage}
    \end{subfigure}
    \par\medskip
    \begin{minipage}{\linewidth}
      \caption{Feature importances across LCS fractions $f$. \psubref{fig:importance-heatmap} Mean standardized MLR coefficient across repetitions for each predictor and buffer scale. Grey cells denote predictors not selected at that fraction and hatched cells predictors selected in only some of the five repetitions; the colour scale is capped at the $f > 0$ range, so the $f = 0$ building-volume coefficient exceeds it. \psubref{fig:importance-headline} GAM term contributions (left) and RF SHAP values (right) for the five predictors with the largest mean standardized MLR coefficient magnitude across the fractions at which LCS stations are present ($f > 0$) in \psubref{fig:importance-heatmap}. Each dot is one station--repetition pair; dot colour encodes the station's feature value (blue = low, red = high). The $f = 0$ column is greyed since only the four AWS contribute to it. Per-model feature importances for all predictors are shown in the section ``Feature importances'' of the results notebook (\nameref{sec:code-lur-results}), and the figure is produced in \nameref{sec:code-lur-importance-figures}.}
      \label{fig:feature-importances}
    \end{minipage}
  \end{adjustwidth}
\end{figure}

Beyond prediction accuracy, the composition of the training network determines which predictors are selected and at which spatial scale (\autoref{fig:feature-importances}).
Since the selection precedes the regression and depends only on the station sample (\nameref{sec:code-lur-model}), all three models share the same predictors at every fraction and repetition, and differ only in how they attribute the response to them.
To limit multicollinearity, a predictor is admitted only if its variance inflation factor against those already selected stays at or below 5 (see \nameref{sec:lur-setup}), so that predictors retained together, even two scales of the same variable, are not redundant with one another.

At $f=0$, only three predictors can enter the model --- with four stations the standardized predictor matrix has rank three --- and in MLR these are building volume at the 250\,m buffer ($+1.33$), tree cover at the 500\,m buffer ($+0.34$) and distance to the lake ($-0.21$) (\autoref{fig:importance-heatmap}).
Notably, two of them, building volume at 250\,m and tree cover at 500\,m, are retained at no other fraction, and tree cover even enters with a warming sign, suggesting that the four-station fit does not capture physically meaningful effects.
As LCS stations are added, the selected feature set broadens markedly, from three predictors at $f=0$ to nine at $f=1$.
Building volume remains the main warming predictor but shifts to the 500\,m scale (selected in three of the five subsamples at $f=0.33$ and in all of them from $f=0.66$, with $+0.80$ at $f=1$), while its 100\,m scale is retained with a much smaller coefficient ($+0.02$ at $f=1$).
Tree cover at 100\,m emerges as a consistent cooling predictor ($-0.27$ at $f=1$), and further topographic predictors (elevation, northness, slope, and TPI at both scales) enter the selection, reflecting the broader range of gradients now represented in the training data.

Most of the predictors selected in only some of the station subsamples enter at $f=0.33$ (hatched cells in \autoref{fig:importance-heatmap}). This is also the fraction at which the predicted temperature range varies most across subsamples (\autoref{fig:pred-spread-range}), so both the selected predictors and the resulting predictions remain sensitive to which stations are included while the network is small.
For the five predictors with the largest MLR coefficient magnitudes (\autoref{fig:importance-headline}), GAM term contributions and RF SHAP values share the sign of the MLR coefficients in all cases: stations with high building volume at the 500\,m scale receive positive contributions (warming), while stations with dense tree cover within 100\,m, on north-facing slopes, at high elevation or far from the lake receive negative contributions (cooling).
The parsimony of the AWS-only selection therefore reflects a sampling artefact rather than physical simplicity, and cannot be read as evidence of which predictors matter: four stations can support only three predictors, and which three are retained, and with which sign, depends on where the stations are located (\nameref{sec:code-nfour-control}).

Together, these results show that network representativeness --- how the station sample covers the city's predictor feature space --- is the primary determinant of reliable city-wide LUR prediction, more so than model class or station density per se.
Model class matters less than coverage: with the four AWS alone, all three models cross-validate well, yet they underestimate the city's temperature range and rest on a predictor set dictated by where the stations happen to be.  %
The models only diverge once coverage broadens, when GAM and RF retain their cross-validation score while that of MLR drops.

\section*{Discussion}

\subsection*{Urban climate monitoring as a focal site study design}

The LUR approach used here is structurally analogous to focal site multi-scale study designs in landscape ecology \cite{brennen2002focal}.
In both frameworks: (i) the aim is to establish a relationship between spatial pattern and an environmental process, (ii) the response consists of measurements at a sample of locations, and (iii) the predictors are spatial metrics computed at multiple buffer scales around each location. In landscape ecology, the choice of focal sites is recognized as critical to the validity of the pattern--process relationship that is ultimately inferred \cite{albert2010sampling,eigenbrod2011sub,pasher2013optimizing,hesselbarth2024computational}.
Our results show that the same holds for urban LUR: where stations are placed determines which parts of the feature space are represented in the training data, and thereby which predictors are selected, with which sign, and how well the fitted model generalizes to the wider city.
Urban climate studies have addressed aspects of network representativeness, particularly through spatially-based station placement optimization \cite{honjo2015network,yang2019designing,malings2018surface,chen2022urban}.
Nevertheless, to our knowledge, the explicit connection to the focal site study framework of landscape ecology --- and the broader methodological challenges highlighted by this analogy --- has not been explored to this extent.

Drawing on this literature, we identify several such challenges that arise equally in urban monitoring networks and the LUR models fitted to them: site selection bias, pseudoreplication from overlapping site buffers, identification of the characteristic spatial scales at which predictors operate, multicollinearity among predictors, and the interpretability--performance trade-off between linear and non-linear models.
The following sections discuss these issues, grounding the analogy in our empirical results and outlining practical implications for LUR modelling and urban climate monitoring network design.

\subsection*{Spatial extrapolation and predictor coverage}

The extrapolation problem in predictive modelling arises when a model predicts for observations whose predictor values lie outside the training range \cite{miller2004spatial,gutzwiller2023using}; in the spatial case, fitting a regression on only a portion of the potential range of spatial predictors can reduce the likelihood of detecting significant effects \cite{brennen2002focal} and lead to contradictory findings in the presence of non-monotonic effects \cite{eigenbrod2011sub}.
Trained on the four AWS alone, all three models underestimate the city's temperature range, notably at its warm end (\autoref{fig:pred-spread}).
Once LCS stations are added, all three models stay below the warm end of the stations, which densely cover the urban core where temperatures are highest, but differ at the cool end: RF stays at the level of the coolest stations, whereas MLR and especially GAM extend beyond it.
This follows from how each model behaves outside its training domain: RF predictions are bounded by the range of the training responses \cite{breiman2001random,zurell2012predicting,lucas2020translucent,meyer2021predicting}, whereas the additive shape functions of GAM can combine into values beyond it \cite{lou2013accurate}.
A similar bias was reported for RF temperature mapping in Zurich, which underestimated high and overestimated low temperatures \cite{zumwald2021mapping}.
Between these two, MLR extends its fitted slopes beyond the sampled range without bound, but its poorer fit to the stations (\autoref{fig:regr-scores}) pulls its predictions towards the mean.
This is also why its warmest predictions stay below those of GAM and RF (\autoref{fig:pred-spread-warm}).

Whether the cooler predictions of GAM and MLR are realistic cannot be validated, since not even the full LCS network samples these environments.
Their higher elevation and distance from the urban core nonetheless make it plausible that they are cooler at night than any station, in which case the observed range is only a lower bound of the city's temperature range, and one that RF cannot exceed by construction.
None of these differences is visible in cross-validation, which does not consider dissimilarities in feature space between the sampled locations and the prediction domain \cite{meyer2021predicting}.

In line with the ``area of applicability'' framework \cite{meyer2021predicting}, the PCA presented here allows diagnosing such coverage gaps before any model is fitted: projecting stations and the target prediction domain into a low-dimensional space makes these gaps immediately apparent.
Previous work in Zurich mapped areas of extrapolation predictor by predictor and found the least reliable predictions in and around wooded areas \cite{zumwald2021mapping}; our PCA-based approach complements this by assessing coverage across all predictors jointly.
In our case, it shows that the AWS miss both ends of the environmental gradient, the forested, elevated periphery and the densely built lakeshore core, which likely correspond to the thermal extremes underpinning the UHI effect.  %
Adding LCS closed the gap in the urban core, but not on the wooded hils around the city.
As a result, the warmest predictions can be checked against stations, whereas the coolest rest on extrapolation.

The two ends of the temperature range do not matter equally in practice.
The warm end matters most for health, as it lies in the densely built core where population density is highest, and the LCS reveal it to be more than 1\,\degree C warmer than the AWS alone suggest (\autoref{fig:pred-spread}).
This is consistent with the finding that reference AWS underestimate urban heat across four Swiss cities, including Zurich \cite{bosch2026revisiting}.
An AWS-only map thus understates heat precisely where it matters most for public health.
The cool end is less critical for health, since the coolest environments are largely forested hillsides where few people live.
Nonetheless, it remains relevant for planning, as vegetated slopes can act as nocturnal cold-air source areas whose drainage ventilates the adjacent neighbourhoods \cite{bigg2014synoptic,grunwald2020predicting}, which the city of Zurich recommends preserving by keeping ventilation corridors in its spatial planning \cite{wymann2012klimaanalyse}.
For network design, this suggests a clear order of priorities: dense LCS networks already capture the conditions that matter most for heat-related health risks, and a third of them suffices to recover the warm end.  %
A few targeted stations in the unsampled cool terrain would then not only validate the coolest predictions, which currently rest on extrapolation, but also help quantify the cooling that these areas provide to the adjacent neighbourhoods and hence inform the measures that seek to preserve it.

Beyond diagnosis, the feature-space coverage view also points towards a concrete optimization framework for future station placement.
Inspired by \citet{bowler2022optimising}, one can discretize the range of each spatial predictor over the regular grid into bins and define a feature-space coverage score as the proportion of occupied bins that are also covered by the station sample.
New station locations could then be selected iteratively to maximize coverage of under-represented bins.
This approach complements existing methods based on spatial clustering \cite{honjo2015network} or urban microclimate simulations \cite{malings2018surface,chen2022urban} by explicitly targeting the representativeness of the predictor feature space rather than spatial proximity alone.

\subsection*{Site buffer overlap need not constrain network design}

A long-standing concern in focal-site studies is that nearby sampling locations can have overlapping landscape buffers, violating assumptions of independence and inducing pseudoreplication \cite{hurlbert1984pseudoreplication,brennen2002focal,eigenbrod2011sub}, which has prompted recommendations to discard proximate sites \cite{holland2004determining,eigenbrod2011sub,pasher2013optimizing}.  %
However, both empirical and simulation studies suggest that overlapping landscapes do not necessarily induce spatial autocorrelation in model residuals and that removing nearby sites can inadvertently constrain the scale of analysis and reduce statistical power to detect predictor effects \cite{zuckerberg2012overlapping,zuckerberg2020review}.

More broadly, residual spatial autocorrelation reflects unaccounted sources of environmental variation, and thus data sampled over a representative region are often preferable to data systematically collected from a spatially restricted one \cite{braunisch2010predicting,zuckerberg2020review}.
For urban air temperature, example sources include anthropogenic heat, surface roughness or synoptic conditions such as wind direction, which can alter the spatial temperature pattern from one night to another.
In urban LUR, buffer overlap is particulary hard to avoid.
Urban environments are highly heterogeneous at the scale of city blocks \cite{collins2000new,cadenasso2007spatial,grimm2008global,zhou2017shifting}, and air temperature can vary markedly over short distances, so capturing intra-urban temperature differences requires closely spaced stations \cite{fenner2017intra,cao2021within}.

Our results support the view that buffer overlap need not undermine inference.
Although buffer overlap grows as LCS are added, residual Moran's $I$ remains close to zero and non-significant for GAM and RF, and is significant only for MLR (\autoref{fig:regr-rsa}).  %
Given that all three models are fitted on the same stations and predictors, and thus with the same buffer overlap, differences in residual spatial autocorrelation point to model misfit rather than overlap.
The low residual autocorrelation of the non-linear models is in line with an independent field campaign in Geneva, whose mobile microclimate transects sample the urban landscape far more extensively than any fixed-station network, yet likewise yielded GAM-based thermal-comfort models with low residual spatial autocorrelation \cite{fahy2025beyond}.
For urban climate network design, this means that stations can be placed to cover the predictor feature space without regard to buffer overlap.
This does not imply, however, that more stations are always better: a third of the LCS network already recovered nearly the full predicted temperature range, which suggests that where stations are placed can matter more than how many there are.

\subsection*{Multi-scale predictors can capture distinct physical processes}

Environmental processes can operate at characteristic scales that are not known a priori and are typically determined empirically \cite{hoek2008review,su2009distance,jackson2012size,llaguno2020environmental}. A common practice is to compute each spatial predictor at multiple buffer radii and retain the scale showing the strongest relationship with the response --- the ``scale of effect'' in ecological studies \cite{jackson2012size,jackson2015ecologists} and the ``scale of influence'' in urban environmental research \cite{llaguno2020environmental}.
Yet the same predictor can mediate multiple processes simultaneously, each potentially operating at a distinct characteristic scale.
Building volume, for instance, can affect air temperature both through radiative exchange, by storing heat and trapping longwave radiation, and through convective exchange, by reducing ventilation, and these processes may act over different extents.
In one urban air quality study, for example, correlations peaked at spatial extents of approximately 150~m for building heights, 300~m for green infrastructure, and 1000~m for parameters governing ventilation or vehicular emissions \cite{llaguno2020environmental}.
Selecting a single scale per predictor therefore risks conflating distinct mechanisms or discarding relevant signal \cite{su2009distance,miguet2016determines}.
Conversely, retaining multiple scales can introduce substantial collinearity, leading to unstable or misleading identification of relevant predictors \cite{graham2003confronting,smith2009confronting,eigenbrod2011sub,dormann2013collinearity,pasher2013optimizing} --- a concern central to the framework adopted here.

To address this challenge, we used a Spearman-ordered sequential forward selection that admits a predictor only if its variance inflation factor against those already selected stays at or below 5 \cite{dormann2013collinearity}, allowing multiple scales of the same predictor to coexist only when they contribute independent information.
Even under this strict criterion, building volume entered at both 100 and 500~m once LCS stations were added (\autoref{fig:importance-heatmap}), two scales that align with physically distinct mechanisms.
The 100~m signal corresponds to the direct radiative influence of the immediate built form, which reduces the sky view factor and traps outgoing longwave radiation \cite{oke1982energetic}, consistent with $\sim$150~m extent reported for building height \cite{llaguno2020environmental}.
The 500~m signal instead arises from the advective nature of air temperature: air exchanges heat with the surfaces it moves over, so the temperature measured at 2~m reflects not only the surface directly beneath the sensor but also an upwind area, i.e., the fetch \cite{oke2007siting,stewart2012local}.
The dominance of the 500~m scale across all models suggests that this upwind heat exchange and neighbourhood-scale ventilation are the primary drivers of near-surface air temperature variability in Zurich.
The contribution of the 100~m scale is much smaller, especially in MLR, but GAM and RF consistently attribute a warming contribution to it (\nameref{sec:code-lur-results}), indicating that the immediate building geometry carries information beyond these neighbourhood-scale processes.  %
This suggests that the two scales are not redundant predictors but proxies for two distinct processes.

The selected scales were themselves sensitive to the composition of the network: with the four AWS alone, building volume entered at 250~m and tree cover at 500~m, neither of which was retained in any station sample including LCS.
With the full network, tree cover was instead retained at 100~m, consistent with the cooling of air temperature by tree canopy being strongest at about the scale of a city block \cite{ziter2019scale,bai2024quantifying}.
Since scales of effect are typically determined empirically, they are only as reliable as the station sample from which they are estimated.
A complementary direction for future work is the use of scalograms, which report the full scale-sensitivity profile of each predictor and could reveal whether the multi-scale effects observed here are consistent across stations and environmental gradients \cite{su2009distance,pasher2013optimizing,ziter2019scale,hesselbarth2024computational}.

\subsection*{Non-linear models can improve robustness but complicate effect attribution}

The three model classes represent different points on a trade-off between predictive performance and interpretability.
When trained exclusively on AWS, all three performed comparably in cross-validation and all underestimated the temperature range of the city, suggesting limitations in station sample rather than in the choice of model.
As LCS data expanded the predictor feature space, GAM and RF retained their cross-validation scores, whereas those of MLR degraded markedly and its residuals became spatially autocorrelated, consistent with a linear form that cannot represent the non-linear, threshold-like relationships between urban form and air temperature \cite{ziter2019scale,bai2024quantifying,hong2025interpreting}.
Two kinds of limitation should therefore be distinguished: a model that cannot fit the sampled conditions, which cross-validation and residual diagnostics reveal, and a station sample that does not cover the city.  %
As discussed above, the latter affects even GAM and RF, which score alike yet extrapolate differently where the network does not sample, with RF bounded by the coolest stations and GAM extending beyond them.
Neither cross-validation nor residual diagnostics can detect this difference.

Beyond predictive performance, the choice of model shapes how results can inform planning.
The spatial predictors used here --- building volume, tree canopy, elevation, topography --- have clear links to the physical processes that regulate urban climate \cite{phelan2015urban} and, unlike remote-sensing-derived indices, can be modified in synthetic scenarios to reflect master plans, densification prospects, or greening interventions \cite{bosch2021evaluating}.
This makes the LUR framework a practical tool for comparing the expected thermal implications of alternative urban design decisions.

Reliable attribution of temperature differences to individual predictors, however, is not straightforward.
In this regard, MLR has a clear advantage in that it offers direct interpretability of coefficients as marginal effects, whereas non-linear models complicate attribution.
Feature importances based on SHAP values, commonly used to interpret RF predictions, reflect statistical associations internal to the model rather than physical mechanisms, and are sensitive to predictor correlation, model architecture, and choice of baseline \cite{hooker2021unrestricted,yu2021study,carloni2025role,campsvalls2025xai} --- concerns that are particularly relevant given the multi-scale, correlated predictors used here.
Shape functions in GAM offer a partial middle ground: their term contributions are exact rather than approximated, though likewise baseline-dependent and sensitive to predictor correlation.

More fundamentally, machine learning models are optimized for predictive performance, and the model that predicts best does not necessarily reflect the causal relationships underlying the data \cite{pichler2023machine,arif2022predictive}.
Given the growing adoption of RF in both landscape ecology \cite{stupariu2022machine} and urban climate research \cite{shi2021monitoring,wang2023machine,hong2025interpreting}, understanding the limitations of common interpretation methods becomes particularly important.
Nevertheless, the warming effect of building volume at the 500\,m scale and the cooling effects of tree canopy, north-facing slopes, elevation and distance from the lake are consistent across all three models, which suggests that these associations are robust to model choice and provides stronger evidence than any single model alone.

Beyond statistical models, effect attribution can be strengthened either through causal inference frameworks, which make the assumed relationships among variables explicit to decide which covariates to account for \cite{zhao2021causal,arif2022utilizing,van2025scaling}, or through physically based models, which represent the underlying processes explicitly.
As we argue next, both depend on representative observations just as much as the regressions studied here.

\subsection*{Network representativeness matters beyond regression}

The coverage gaps identified here reflect a broader mismatch between the spatial complexity of urban thermal environments and the data and models used to represent it.  %
Causal inference frameworks require positivity, i.e., that the covariate combinations of interest are represented in the sample \cite{petersen2012diagnosing}, a condition that AWS alone fail to meet given their coverage gaps in predictor feature space.
In spatial settings, such violations force effects to be extrapolated beyond the sampled conditions \cite{gilbert2021causal}, which can bias results without apparent signs of failure, as in the underestimated city-wide temperature range of the AWS-only models reported here.

Physically based models span a wide range of scales, from urban land surface schemes that simulate the surface energy balance to mesoscale atmospheric models and building-resolving large-eddy simulation models.
Despite their increasing level of detail, these models show large variations in skill and are rarely evaluated against dense observation networks.
In the Urban-PLUMBER project, for instance, thirty urban land surface schemes evaluated at a single suburban flux site differed widely in performance, with the most complex ones generally not outperforming simpler ones \cite{lipson2024evaluation}.
In one of the few intercomparisons against a dense network of 70 LCS in Bern, Switzerland, shortcomings were revealed in all three numerical models evaluated, which either underestimated intra-urban temperature variability, overestimated small-scale gradients or showed a large warm bias, and were all outperformed by a LUR calibrated on the same stations \cite{burger2024spoilt}.
This underestimation of intra-urban variability, reminiscent of the AWS-only regressions studied here, was detectable because the evaluation network spanned a wide range of urban conditions.
Against AWS alone, which tend to sample a much narrower range, it would have been considerably harder to detect.

The same holds for the ecosystem service models increasingly used to inform urban greening, of which the InVEST urban cooling model is a widely applied example \cite{ronchi2020urban,bosch2021spatially,hamel2024calibrating,silveira2024importance,huang2025quantifying,nabikandi2025evaluating}.
Its simulated temperatures are bounded between a rural reference temperature and a maximum UHI magnitude, often derived from the coolest and warmest stations available \cite{ronchi2020urban,silveira2024importance,ramsay2025assessing}, so that a network that misses the warmest locations compresses the entire simulated range.
Even with 27 LCS in Singapore, the maximum UHI was likely underestimated because most sensors were in vegetated areas, which led \citet{ramsay2025assessing} to suggest designing networks to capture the densely built-up areas where the UHI peaks.
In Zurich, the LCS close precisely this gap that the AWS leave at the warm end, but in turn leave the forested, cool terrain unmonitored.
Both are instances of the same problem: networks that cover only a part of the urban environmental gradient, leaving the rest of the city without observations to fit or evaluate any model.

This also bears on a question raised by \citet{burger2024spoilt}, namely how many stations a reliable regression requires.
Our results suggest that the answer depends on where the stations are as much as on how many there are.
A third of the LCS network already recovered nearly the full predicted temperature range, whereas the predictors selected, and their signs, depended on how the stations sampled the city's feature space.

Network representativeness is thus neither a matter of station count alone nor specific to regression.
An unrepresentative station sample can bias causal effect estimates and the evaluation of physically based models, just as it biases the spatial extrapolation of a regression.
The feature-space coverage framework developed here can be applied to any use of a station network, whether to train a regression, to estimate causal effects or to calibrate and evaluate a physically based model.
It thus offers a principled basis for designing networks that span the full range of urban conditions, including the extremes that matter most for heat exposure and urban planning.

\section*{Materials and methods}

\subsection*{Study area}

The study area encompasses the urban agglomeration of Zurich, Switzerland, on the Swiss Plateau, which is characterized by a temperate oceanic climate. The extent is defined as the convex hull of all LCS station locations, buffered by 2\,km to include the urban--rural transition zone (\nameref{sec:code-spatial-extent}), covering a total area of 161\,km$^2$.

\subsection*{Air temperature data}
\label{sec:air-temperature-data}

We use air temperature measurements from 121 stations measuring at screen level (2\,m above ground), namely four reference automatic weather stations (AWS) and 117 low-cost temperature sensors (LCS).
The AWS comprise two stations operated by the Federal Office of Meteorology and Climatology (MeteoSwiss) and two of the National Air Pollution Monitoring Network (NABEL).
The LCS comprise 93 sensors of the urban climate network of the city of Zurich (UGZ) \cite{ugz2026stadtklima} and 24 sensors of the air quality monitoring network of the canton of Zurich (AWEL) \cite{awel2026lufttemperatur}.
Of the 25 AWEL sensors within the study area, one is co-located with the NABEL AWS at Zurich-Kaserne and therefore shares its exact coordinates and predictor values, so it is excluded to avoid distorting the nearest-neighbour weights used in the residual spatial autocorrelation diagnostics.
Another AWEL sensor was relocated during the study period, and only its measurements from the location within the study area are used.

We extracted measurements for all days on which the daily-mean air temperature at the warmest AWS exceeded 25\,\degree C during the summers (June--August) of 2022--2025, yielding 78 days grouped into 31 hot-spell periods\footnote{According to the MeteoSwiss heat warning system \cite{burgstall2021von,ragettli2023explorative}, one or two such days correspond to a level 2 warning (hot spell); three or more days constitute a heat wave.}.
The retrieval of the MeteoSwiss measurements, as well as the detection of hot spell periods across all sources, were performed using the meteora Python library \cite{bosch2026meteora}, whereas the AWEL measurements were retrieved directly from the monthly files published by AWEL, which record the location of each sensor at the time of each measurement.

The UHI in Switzerland has been reported to peak around 21:00 local summer time (UTC+2) \cite{burgstall2019representing}.
Since LCS, and particularly the AWEL sensors, are prone to radiative errors \cite{awel2026lufttemperatur,bosch2026revisiting}, which respond to solar radiation with a lag of one to two hours \cite{cornes2020correcting,beele2022quality,bosch2026revisiting}, we use the 23:00 observation, which follows sunset by at least one and a half hours on all selected days.

\begin{figure}[ht]
  \begin{adjustwidth}{-.4\textwidth}{0cm}
    \includegraphics[width=\linewidth]{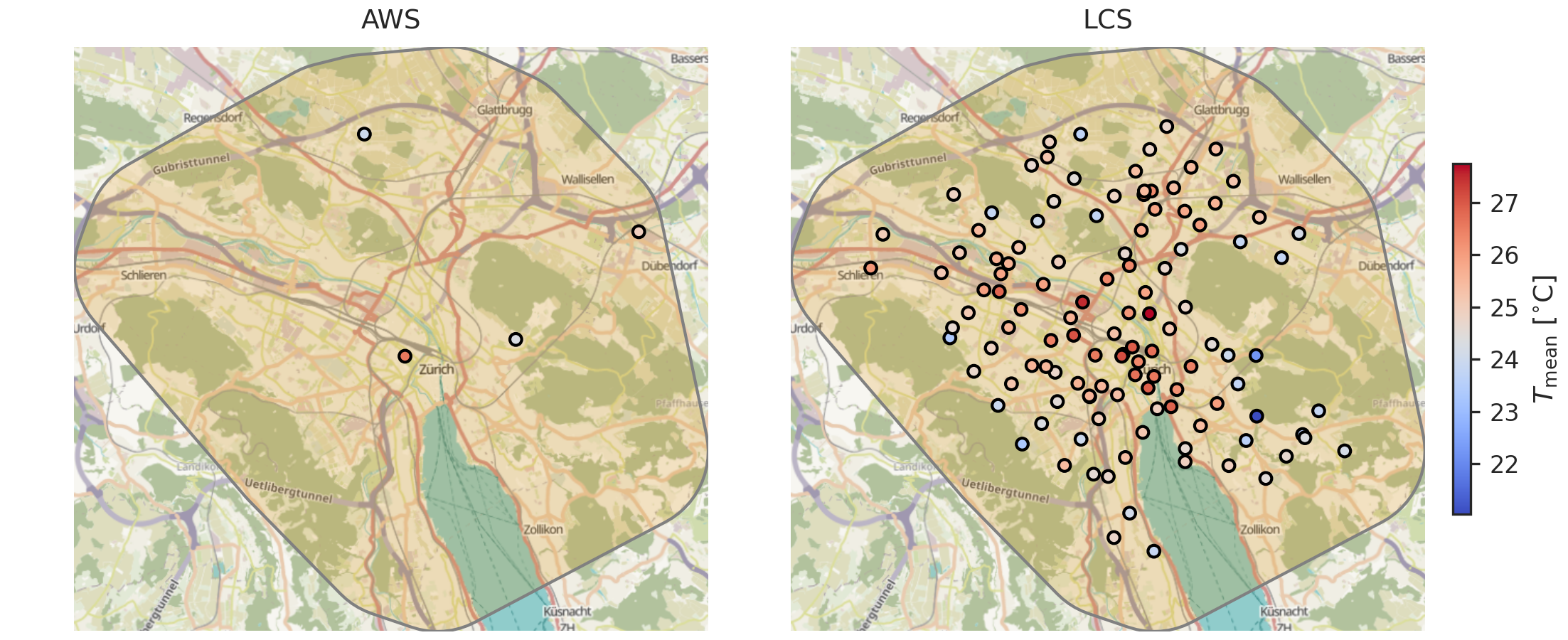}
    \begin{minipage}{\linewidth}
      \caption{AWS (left) and LCS (right) station locations in Zurich, coloured by mean air temperature $T$ over all hours of the selected days. \label{fig:station-temperatures}}
    \end{minipage}
  \end{adjustwidth}
\end{figure}

\subsection*{Spatial predictor variables}

For each station and grid cell, we compute a set of geographic predictors derived from spatial data (\autoref{tab:spatial-predictors}). Building volume is estimated by combining building footprints from OpenStreetMap (OSM) with LiDAR data from swissSURFACE3D \cite{swisstopo2024swisssurface3d}. Tree canopy proportion is derived from the same LiDAR product by filtering high-vegetation points and rasterizing at 1\,m resolution. Elevation is extracted from the swissALTI3D digital elevation model (DEM) at 2\,m resolution \cite{swisstopo2024swissalti3d}, from which we also derive slope and aspect using the xDEM library \cite{xdem2024}, as well as the topographic position index (TPI).
While elevation is the absolute height above sea level at the site, TPI measures relative topographic position, i.e., the difference between the elevation at the site and the mean elevation over the surrounding neighbourhood \cite{weiss2001topographic}, so that positive values denote local convexities (crests, ridges, spurs) and negative values local concavities (valleys, hollows), independently of the absolute altitude.
Distance to Lake Zurich is added to account for lake--land thermal interactions, including lake and land breezes, which in complex terrain interact with local circulations and urban heat \cite{brandi2025urbanization}.
We use building volume rather than a built-up or impervious-surface fraction, since it captures the three-dimensional built form, i.e., footprint, height and hence canyon geometry, that a two-dimensional cover fraction omits.

All predictors except elevation and lake distance are computed within concentric circular buffers. Building volume and tree canopy use radii of 100, 250 and 500\,m; TPI, slope and northness use 250 and 500\,m.  %
Topographic predictors are not computed at 100\,m, which is too small to represent the upwind area that shapes near-surface air temperature \cite{oke2007siting,stewart2012local}.
For slope and aspect, the attribute is computed per DEM cell and then averaged within the buffer, so the radius sets the extent over which the attribute is aggregated; for TPI, the radius instead defines the neighbourhood against which the site elevation is compared, and hence the scale of the attribute itself. Multi-scale predictors are computed using the focalpy library \cite{bosch2026focalpy}.

\begin{table}
  \setlength{\tabcolsep}{9pt}
  \begin{adjustwidth}{-.4\textwidth}{0cm}
    \raggedright
    \begin{tabular}{l l l l}
      \toprule
      \textbf{Feature} & \textbf{Description} & \textbf{Scales} & \textbf{Data source} \\
      \midrule
      Building volume & Total building volume & 100, 250, 500 m & OSM, swissSURFACE3D \\
      Tree canopy & Proportion of tree canopy & 100, 250, 500 m & swissSURFACE3D \\
      Elevation & Elevation at the site location & -- & swissALTI3D \\
      Slope & Mean elevation gradient & 250, 500 m & swissALTI3D \\
      Northness & Mean cosine of aspect & 250, 500 m & swissALTI3D \\
      TPI & Site elevation relative to neighbourhood mean & 250, 500 m & swissALTI3D \\
      Lake distance & Distance to Lake Zurich & -- & OSM \\
      \bottomrule
    \end{tabular}
    \caption{Spatial predictor variables used in the land use regression.}
    \label{tab:spatial-predictors}
  \end{adjustwidth}
\end{table}

\subsection*{Land use regression set up}
\label{sec:lur-setup}

The target variable for each station is the temperature anomaly $T - T_{\mathrm{mean}}$, i.e., the difference between the station's temperature and the mean of all concurrent station measurements used to fit the model, so that each station contributes one observation per selected day.
This removes the synoptic signal and isolates the spatial component of intra-urban temperature variability.
Following \citet{burger2024spoilt}, we reference the anomaly to the mean of all concurrent stations rather than to a single rural station, as in previous LUR studies of urban air temperature \cite{burger2021modelling,burger2022modeling,oukawa2022fine}, so that the target is not tied to the measurement errors and data gaps of any single station.
As most stations are urban, this mean is itself affected by the UHI, so the anomalies express temperature differences within the network rather than urban--rural contrasts.
The choice of reference does not affect the spread of the anomaly across stations, and hence the UHI magnitude, because a common reference shifts all concurrent measurements equally.

We compare three models: multiple linear regression (MLR), a generalized additive model (GAM) implemented as an explainable boosting machine (EBM) \cite{lou2013accurate} without interactions via the InterpretML library \citep{nori2019interpretml}, and random forest (RF) with data-dependent default hyperparameters from FLAML's zero-shot auto machine learning \citep{wang2021flaml,kayali2022mining}. Each model is wrapped in a scikit-learn pipeline \cite{pedregosa2011scikit} which is composed of feature standardization, a Spearman-ordered sequential forward selection and the model.
The feature selector ranks features by the absolute Spearman correlation with the target and adds them greedily.
A candidate is admitted only if its variance inflation factor against the already selected features stays at or below 5, as in \citet{alonso2020new}, a common safeguard against multicollinearity \cite{dormann2013collinearity,shi2018modelling,shi2019assessing,alonso2020new,oukawa2022fine}.
Since a later candidate can raise the variance inflation factor of features selected before it, a backward pass then repeatedly removes the selected feature with the highest variance inflation factor until all of them stay at or below 5.
This selection generalizes the common LUR practice of retaining, for each predictor, the buffer radius best correlated with the target \cite{su2009distance,zhou20142011,shi2018modelling,burger2021modelling,alonso2020new}, i.e., its scale of effect, while letting several scales of a predictor enter the model as long as they are not collinear.
With four stations the standardized predictor matrix has rank three, so any fourth candidate is an exact linear combination of the others and its variance inflation factor is infinite, which caps the AWS-only selection at three predictors.

We use the four-AWS configuration as a baseline and progressively augment it with 39, 77 and all 117 LCS at $f = 0.33$, $0.66$ and $1$, respectively.
For intermediate fractions, stations are subsampled with five repetitions to assess sensitivity to the specific subset; the full-LCS case is deterministic.
In line with previous spatiotemporal LUR studies \cite{shi2018modelling,oukawa2022fine}, model performance is evaluated using three-fold cross-validation with the daily observations of all stations split at random into folds.
From the out-of-fold predictions, we compute $R^2$, RMSE and MAE, and we assess the spatial structure of the errors \cite{alonso2020new} with Moran's $I$ \citep{moran1950notes} of the residuals averaged per station, using $k=4$ nearest-neighbour weights and 999 permutations ($p_{\mathrm{sim}}$) in PySAL \citep{rey2007pysal}.  %
Since the observations of a station on different days can fall in different folds, these scores measure how well the models reproduce the conditions at the station locations rather than elsewhere in the city.
Holding out whole stations instead \cite{roberts2017cross} is hardly feasible with four AWS, as each fold would then be trained on only two or three stations.

Feature importance is extracted as standardized coefficients for MLR and as per-station SHAP values \citep{lundberg2017unified} for GAM and RF, in line with previous machine learning models of urban air temperature \cite{oukawa2022fine,hong2025interpreting}.
For GAM, contributions are obtained with SHAP's additive explainer, which is exact for an interaction-free EBM. For RF, SHAP values are computed using TreeExplainer \citep{lundberg2020local} (\nameref{sec:code-lur-model}).
The SHAP values are shown as beeswarm plots, with dot colour encoding the station's feature value.

\subsection*{Spatial prediction and coverage assessment}

To produce city-wide thermal maps, we compute the same spatial predictors on a regular 100\,m grid over the study area using the focalpy library \cite{bosch2026focalpy}, and apply each fitted model to all grid cells, averaging the predictions over the subsampling repetitions (\nameref{sec:code-lur-model}).
Grid cells on Lake Zurich (8\,km$^2$) are excluded, since their land use predictors are degenerate and a land use regression does not apply to air temperature over open water.
Station coverage in feature space is assessed with a PCA of the standardized predictors over the grid cells, computed with scikit-learn \cite{pedregosa2011scikit} (\nameref{sec:code-lur-results}), onto which the stations are then projected  (\nameref{sec:code-lur-results}).

To summarize each prediction surface, we take the range between the 5th and 95th percentiles of the predicted anomaly across grid cells, a robust counterpart of the difference between the warmest and coolest locations that avoids the sensitivity of the minimum--maximum range to individual cells.
For the intermediate fractions, the percentiles are computed on each repetition's own surface and then averaged, since averaging the surfaces first would shrink their extremes.
Since the zero point of the anomaly depends on the stations of each draw, the cool and warm ends of the range (Fig.~\ref{fig:pred-spread}) are expressed relative to the mean of the full network.
Reported intervals are bootstrap 95\% confidence intervals across repetitions.

\section*{Data and code availability}

All data used in this study are openly available.
Air temperature measurements were obtained from MeteoSwiss, the National Air Pollution Monitoring Network (NABEL), the urban climate network of the city of Zurich \cite{ugz2026stadtklima} and the air quality monitoring network of the canton of Zurich \cite{awel2026lufttemperatur}, and the spatial predictors were derived from swissSURFACE3D \cite{swisstopo2024swisssurface3d}, swissALTI3D \cite{swisstopo2024swissalti3d} and OpenStreetMap.
The full analysis, from data retrieval to the figures, is implemented as a reproducible Snakemake pipeline \cite{molder2025sustainable} of Jupyter notebooks, available at \href{https://github.com/martibosch/zurich-uhi-lur}{github.com/martibosch/zurich-uhi-lur} and archived together with the executed notebooks and all generated figures at Zenodo (\href{https://doi.org/10.5281/zenodo.23527239}{doi:10.5281/zenodo.23527239}).
Since the AWEL measurement files are periodically updated upstream, the AWEL data were cached at retrieval time, and this snapshot is included in the archive.

\section*{Supplementary Material}

\setcounter{figure}{0}
\renewcommand{\thefigure}{S\arabic{figure}}

\subsection*{Code S1}
\label{sec:code-lur-results}
Visualization of land use regression results (figures), as a Jupyter notebook (IPYNB).
\href{https://github.com/martibosch/zurich-uhi-lur/blob/arxiv-v1/notebooks/output/lur-results.ipynb}{github.com/martibosch/zurich-uhi-lur/blob/arxiv-v1/notebooks/output/lur-results.ipynb}

\subsection*{Code S2}
\label{sec:code-lur-model}
Model fitting experiment (MLR, GAM and RF across LCS fractions) and city-wide grid prediction, as a Jupyter notebook (IPYNB).
\href{https://github.com/martibosch/zurich-uhi-lur/blob/arxiv-v1/notebooks/output/lur-model.ipynb}{github.com/martibosch/zurich-uhi-lur/blob/arxiv-v1/notebooks/output/lur-model.ipynb}

\subsection*{Code S3}
\label{sec:code-nfour-control}
Four-station control: predictor selection and dominant predictor across random four-station subsets of the LCS network, as a Jupyter notebook (IPYNB).
\href{https://github.com/martibosch/zurich-uhi-lur/blob/arxiv-v1/notebooks/output/lur-nfour-control.ipynb}{github.com/martibosch/zurich-uhi-lur/blob/arxiv-v1/notebooks/output/lur-nfour-control.ipynb}

\subsection*{Code S4}
\label{sec:code-lur-importance-figures}
Feature importance figure: coefficient heatmap across LCS fractions and GAM and RF contributions of the headline predictors, as a Jupyter notebook (IPYNB).
\href{https://github.com/martibosch/zurich-uhi-lur/blob/arxiv-v1/notebooks/output/lur-importance-figures.ipynb}{github.com/martibosch/zurich-uhi-lur/blob/arxiv-v1/notebooks/output/lur-importance-figures.ipynb}

\subsection*{Code S5}
\label{sec:code-spatial-extent}
Study area delineation, as a Jupyter notebook (IPYNB).
\href{https://github.com/martibosch/zurich-uhi-lur/blob/arxiv-v1/notebooks/output/lcs-spatial-extent.ipynb}{github.com/martibosch/zurich-uhi-lur/blob/arxiv-v1/notebooks/output/lcs-spatial-extent.ipynb}

\bibliography{references}
\bibliographystyle{unsrtnat}

\end{document}